\documentclass[twocolumn]{aastex701}

\begin{document}

\title{A post-perihelion constraint on the CO$_{2}$/H$_{2}$O ratio of interstellar comet 3I/ATLAS from [O I] forbidden lines}

\author[orcid=0000-0003-4490-9307,gname=Yoshiharu, sname=Shinnaka]{Yoshiharu Shinnaka}
\affiliation{Koyama Space Science Institute, Kyoto Sangyo University, Motoyama, Kamigamo, Kita-ku, Kyoto 603-8555, Japan}
\email[show]{yoshiharu.shinnaka@cc.kyoto-su.ac.jp}  

\author[orcid=0009-0007-6258-3351,gname=Ko, sname=Tsujimoto]{Ko Tsujimoto} 
\affiliation{Department of Physics, Faculty of Science, Kyoto Sangyo University, Motoyama, Kamigamo, Kita-ku, Kyoto 603-8555, Japan}
\email{i2585075@cc.kyoto-su.ac.jp}

\author[orcid=0000-0003-2011-9159,gname=Hideyo,sname=Kawakita]{Hideyo Kawakita}
\affiliation{Koyama Space Science Institute, Kyoto Sangyo University, Motoyama, Kamigamo, Kita-ku, Kyoto 603-8555, Japan}
\affiliation{Department of Physics, Faculty of Science, Kyoto Sangyo University, Motoyama, Kamigamo, Kita-ku, Kyoto 603-8555, Japan}
\email{kawakthd@cc.kyoto-su.ac.jp}

\author[orcid=0000-0002-1632-5489,gname=Hitomi,sname=Kobayashi]{Hitomi Kobayashi}
\affiliation{Photocross Co., Ltd., Bldg. D, Iwakura-Chuzaichi-cho, Sakyo-ku, Kyoto 606-0021, Japan}
\affiliation{Koyama Space Science Institute, Kyoto Sangyo University, Motoyama, Kamigamo, Kita-ku, Kyoto 603-8555, Japan}
\email{h_kobayashi@photo-cross.com}

\author[orcid=0000-0003-4391-4446,gname=Jun-ichi,sname=Watanabe]{Jun-ichi Watanabe}
\affiliation{National Astronomical Observatory of Japan, 2-21-1 Osawa, Mitaka, Tokyo 181-8588, Japan}
\affiliation{Koyama Space Science Institute, Kyoto Sangyo University, Motoyama, Kamigamo, Kita-ku, Kyoto 603-8555, Japan}
\email{jun.watanabe@nao.ac.jp}

\author[orcid=0000-0002-5413-3680,gname=Takafumi,sname=Ootsubo]{Takafumi Ootsubo}
\affiliation{Department of Basic Sciences, University of Occupational and Environmental Health, Japan, 1-1, Iseigaoka, Yahatanishi-ku Kitakyushu, Fukuoka, 807-8555, Japan}
\email{ootsubo23744@med.uoeh-u.ac.jp}

\begin{abstract}

We present high-resolution optical spectroscopy of interstellar comet 3I/ATLAS (C/2025 N1) obtained with the High Dispersion Spectrograph mounted on the Subaru Telescope on UT 2026 January 7, when the comet was on its outbound trajectory at a heliocentric distance of $r_{\mathrm{h}} = 2.87$ au. The spectra cover the forbidden atomic oxygen lines, [O~I], at 557.7, 630.0, and 636.4 nm. The [O~I] red-doublet intensity ratio $I_{630.0}/I_{636.4} = 2.91 \pm 0.21$ matches the optically thin branching ratio ($\sim$3; \citealt{StoreyZeippen2000}), indicating that optical-depth effects are small and that our relative flux calibration is reliable. We measure a green-to-red [O~I] intensity ratio of $G/R = I_{557.7}/(I_{630.0} + I_{636.4}) = 0.339 \pm 0.027$. This value is higher than those of most Solar System comets at similar heliocentric distances, but comparable to that of the interstellar comet 2I/Borisov. From the measured $G/R$ ratio in 3I/ATLAS, we estimate the CO$_2$/H$_2$O abundance ratio under the assumption that H$_2$O and CO$_2$ are the dominant parents of O($^1$S) and O($^1$D), with other oxygen-bearing species expected to have a smaller influence under typical conditions (e.g., \citealt{FestouFeldman1981}). The derived ratio is significantly lower than the extremely CO$_2$-rich composition reported from infrared observations on the inbound trajectory at $r_{\mathrm{h}} \sim 3.3$ au, yet higher than typical values measured for comets in the Solar System. Together with published pre- and post-perihelion measurements, our result indicates that the CO$_2$/H$_2$O ratio decreased substantially across perihelion.

\end{abstract}




\section{Introduction} 

Interstellar comet 3I/ATLAS (C/2025 N1; hereafter 3I) is the third confirmed interstellar object. It provides a rare opportunity to investigate the physical and chemical properties of icy planetesimals that formed in a protoplanetary disk beyond the Solar System. Since its discovery on UT 2025 July 1 by the Asteroid Terrestrial-impact Last Alert System (ATLAS; \citealt{Denneau2025}), follow-up observations have rapidly established the presence of cometary activity and characterized the dust coma, including constraints on the nucleus size and dust distribution (\citealt{Frincke2026} and references therein).

Chemical characterization of 3I has relied on multiwavelength detections of volatiles, metals, and ice features. Shortly after discovery, no gas emission was detected, but water ice was directly identified in near-infrared spectra obtained with NASA IRTF/SpeX at heliocentric distance $r_{\mathrm{h}} = 4.1$ au \citep{Yang2025}. As the comet approached the Sun, various emission lines began to appear, with Ni~I detected at $r_{\mathrm{h}} = 3.78$ au and CN at $r_{\mathrm{h}} = 3.6$ au \citep{Opitom2025,SalazarManzano2025,Rahatgaonkar2025} and references therein. On its inbound trajectory, space-based infrared observations with JWST and SPHEREx were used to detect gaseous H$_2$O, CO$_2$, and CO in addition to water ice \citep[e.g.,][]{Cordiner2025,Lisse2025}. Taken together, these infrared measurements suggest that 3I was highly CO$_2$-enriched prior to perihelion, with CO also enhanced relative to H$_2$O, compared with typical Solar System comets \citep[e.g.,][]{Ootsubo2012}.

\begin{deluxetable*}{lccccccc}[t]
\tablecaption{Observational circumstances in 2026 January 7\label{tab:obs}}
\tablehead{
\colhead{UT Time} &
\colhead{Object} &
\colhead{$T_{\mathrm{exp}}$} &
\colhead{$r_{\rm h}$} &
\colhead{$\Delta$} &
\colhead{$\dot{\Delta}$} &
\colhead{PA$_{\mathrm{slit}}$} &
\colhead{Airmass} \\
\colhead{} &
\colhead{} &
\colhead{(s)} &
\colhead{(au)} &
\colhead{(au)} &
\colhead{(km s$^{-1}$)} &
\colhead{($^{\circ}$)} &
\colhead{}
}
\startdata
11:15 & HR 3454  &   30  & ---              & ---              & ---            & ---        & 1.06 \\
11:21 & 3I/ATLAS & 1500  & 2.8682--2.8687   & 1.9748--1.9751   & 32.55--32.63   & 161--164   & 1.03 \\
11:47 & 3I/ATLAS & 1500  & 2.8687--2.8693   & 1.9751--1.9754   & 32.63--32.70   & 164--184   & 1.01 \\
12:19 & 3I/ATLAS & 1500  & 2.8694--2.8700   & 1.9755--1.9758   & 32.73--32.80   & 195--249   & 1.00 \\
12:46 & HR 3454  &   30  & ---              & ---              & ---            & ---        & 1.07 \\
13:12 & HR 4963  &   30  & ---              & ---              & ---            & ---        & 1.68 \\
13:14 & HR 4963  &   30  & ---              & ---              & ---            & ---        & 1.67 \\
15:52 & HR 4963  &   30  & ---              & ---              & ---            & ---        & 1.12 \\
15:52 & HR 4963  &   30  & ---              & ---              & ---            & ---        & 1.12 \\
\enddata
\tablecomments{UT Time is the start time of each exposure. $T_{\mathrm{exp}}$ is the exposure time. $r_{\mathrm{h}}$ and $\Delta$ are the heliocentric and geocentric distances, respectively. $\dot{\Delta}$ is the relative velocity with respect to the Earth. PA$_{\mathrm{slit}}$ is the position angle of the slit.}
\end{deluxetable*}

To interpret these infrared constraints, \citet{Maggiolo2026} proposed that long-term galactic cosmic ray processing can create CO$_2$-enriched near-surface layers. They argued that the shallow pre-perihelion erosion expected for 3I would mainly sample this processed material, whereas less-processed layers could contribute only under limited post-perihelion conditions. This scenario provides a potential framework for interpreting the infrared constraints.

Post-perihelion results remain sparse and in some cases discrepant, possibly indicating a decrease in the CO$_2$/H$_2$O ratio from pre- to post-perihelion \citep{Lisse2026}. These uncertainties motivate independent constraints on the relative roles of H$_2$O and CO$_2$ in 3I's coma. Consistent with such evolution, optical narrowband filter imaging suggests that 3I transitioned from carbon-chain depleted before perihelion to carbon-normal after perihelion \citep{Jehin2025}.

In addition to direct measurements of H$_2$O and CO$_2$ with space-based infrared facilities, the forbidden emission lines of atomic oxygen (hereafter [O~I]) at 557.7, 630.0, and 636.4 nm provide an alternative pathway to infer the CO$_2$/H$_2$O abundance ratio in comets \citep[e.g.,][]{Furusho2006,Decock2013,Decock2015,McKay2015}. These lines are emitted via radiative decay of metastable oxygen atoms, O($^1$S) and O($^1$D), which are produced directly by photodissociation of oxygen-bearing parent molecules. O($^1$S) emits the green line at 557.7 nm primarily via radiative decay to O($^1$D). An alternative decay produces the UV doublet at 297.2 and 295.8 nm, but its branching ratio is small ($\sim$5\%). Therefore, detection of the green line necessarily implies the presence of the red-doublet emission at 630.0 and 636.4 nm from subsequent radiative decay of O($^1$D) to the ground O($^3$P) states, exept for the small fraction that decays via the UV doublet. In Solar System comets, the dominant parents are H$_2$O and CO$_2$ \citep{Biver2024}. Consequently, the relative intensities of the green line ([O~I] 557.7 nm) and the red doublet ([O~I] 630.0/636.4 nm), expressed as the $G/R$ ratio, are sensitive to the mix of parent molecules and can be used to constrain the CO$_2$ contribution relative to H$_2$O \citep[e.g.,][]{FestouFeldman1981}. At the same time, the observed [O~I] ratios can be influenced by collisional quenching of O($^1$D) and O($^1$S) in the inner coma (typically within $\sim$1000 km from the nucleus at $r_{\mathrm{h}} = 1$ au), as well as by observing geometry and aperture effects.

Here we present high-dispersion optical observations of 3I/ATLAS obtained on UT 2026 January 7, when the comet was on its outbound trajectory at $r_{\mathrm{h}} = 2.87$ au. We analyze the [O~I] forbidden lines to constrain the relative contributions of H$_2$O and CO$_2$ to atomic oxygen production in this interstellar comet. Our measurements provide an opportunity to compare the CO$_2$/H$_2$O ratio before and after perihelion, which is important for assessing compositional evolution through perihelion. Our results also enable a direct comparison with the established [O~I] framework for Solar System comets.

\section{Observations And Data Reduction}

We conducted high-resolution optical spectroscopic observations of the third confirmed interstellar object, 3I/ATLAS, on UT 2026 January 7, after its perihelion passage on UT 2025 October 29, using the High Dispersion Spectrograph (HDS; \citealt{Noguchi2002}) mounted on the Subaru Telescope at Maunakea, Hawaii. At the time of the observations, the heliocentric and geocentric distances of the comet were $r_{\mathrm{h}} = 2.87$ au and $\Delta = 1.98$ au, respectively. We used a $0\farcs5 \times 9\farcs6$ slit, providing a resolving power of $R = 72{,}000$ (corresponding to $\sim$4.2 km s$^{-1}$) over the wavelength range 553--810 nm, with a gap between 673 and 697 nm. This setup covers the three [O~I] lines at 557.7, 630.0, and 636.4 nm. The slit length corresponds to a projected tangential distance of $\pm 6892$ km from the nucleus. The optical center of the comet was kept on the slit throughout the observations, resulting in a total on-source integration time of 4500 s ($1500$ s $\times$ 3 frames). The slit orientation was mechanically fixed, and therefore its position angle on the sky varied from $161^\circ$ to $249^\circ$ during the observations. The position angle of the Sun--comet radius vector and the solar phase angle (Sun--comet--observer) were $290\fdg6$ and $10\fdg1$, respectively. Details of the observations are summarized in Table~\ref{tab:obs}.

\begin{figure}[t]
\centering
\includegraphics[angle=-90,width=0.48\textwidth]{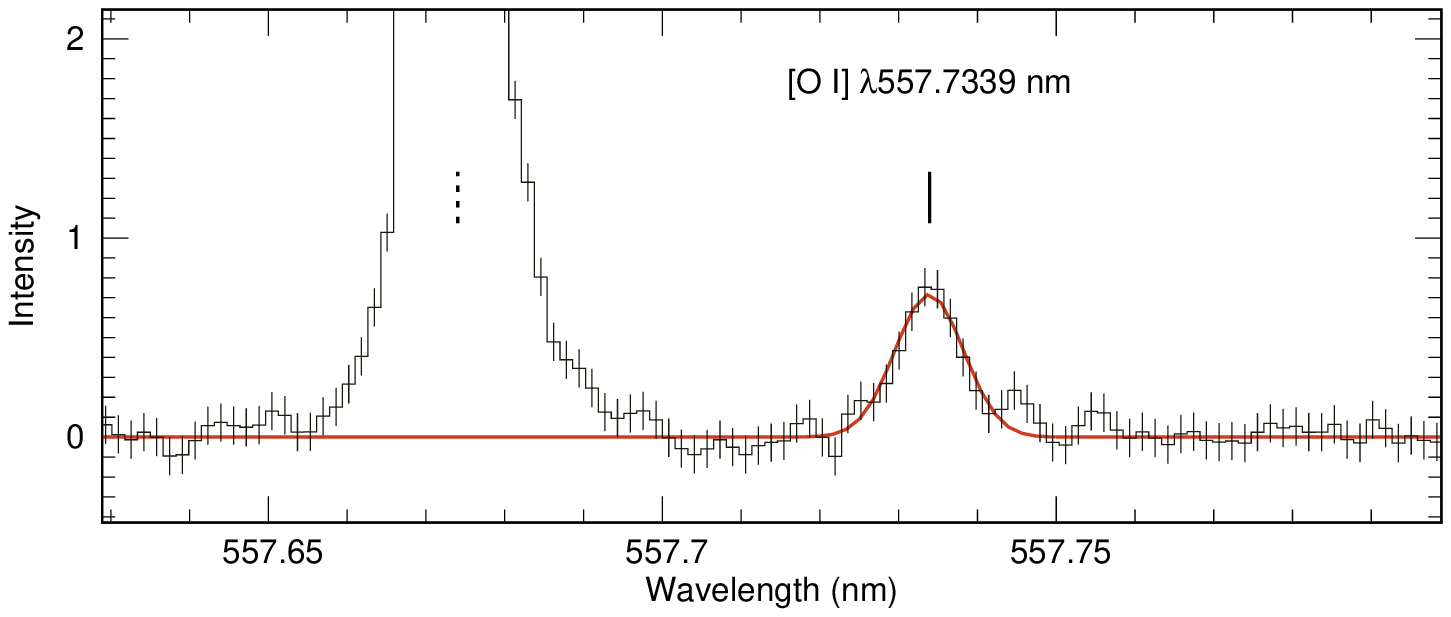}\par\vspace{0.6ex}
\includegraphics[angle=-90,width=0.48\textwidth]{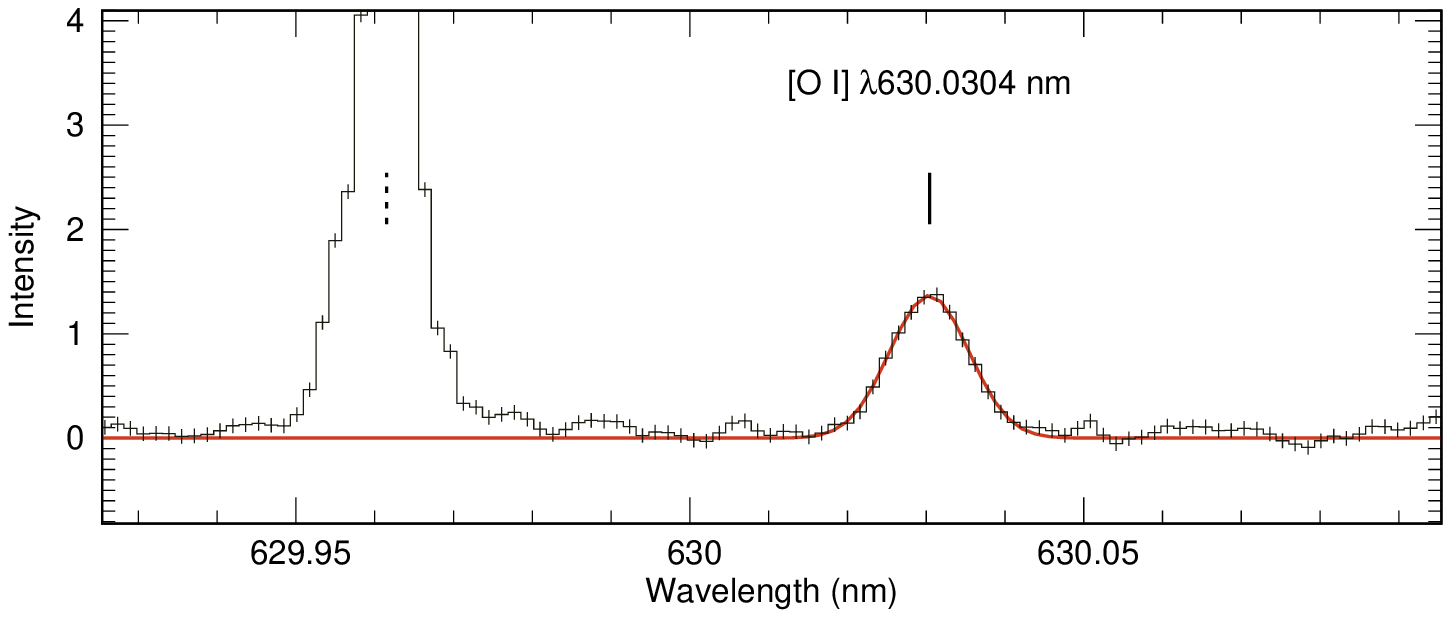}\par\vspace{0.6ex}
\includegraphics[angle=-90,width=0.48\textwidth]{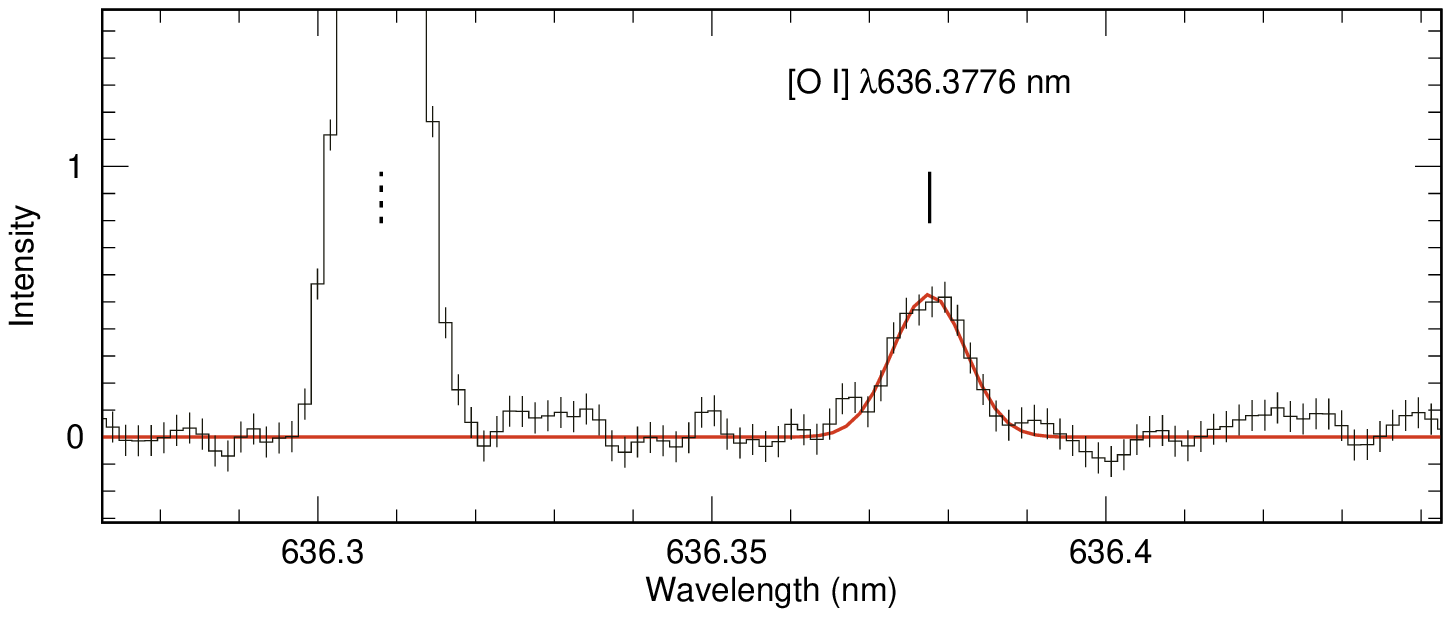}
\caption{Spectra of the three [O~I] emission lines around 557.73 nm (top), 630.03 nm (middle), and 636.37 nm (bottom) in 3I/ATLAS observed on UT 2026 January 7. Wavelengths are given in the comet rest frame. Our best-fit single-Gaussian profiles for the three [O~I] lines are overplotted as red curves. Vertical ticks indicate the cometary [O~I] lines, and vertical dotted lines denote the positions of the telluric [O~I] lines. The telluric oxygen emission lines are clearly separated from the cometary lines. We neglect possible contamination of the [O~I] lines by other species (C$_2$ and NH$_2$), because their expected intensities are below the measurement uncertainty.}
\label{fig:OI_lines}
\end{figure}

The data were reduced using the IRAF software package\footnote{IRAF is distributed by the National Optical Astronomy Observatory, which is operated by the Association of Universities for Research in Astronomy (AURA) under cooperative agreement with the National Science Foundation.} following the standard HDS reduction procedures\footnote{\url{http://www.naoj.org/Observing/Instruments/HDS/hdsql-e.html}}, including bias subtraction, linearity correction, cosmic-ray rejection, scattered-light subtraction, cross-talk subtraction, and flat-fielding. One-dimensional spectra were extracted by summing the signal over a $9\farcs6$ aperture along the slit centered on the comet. Wavelength calibration was performed using Th--Ar lamp spectra. Flux calibration was carried out using spectrophotometric standard stars (HR 3454 and HR 4963) observed before and after the comet observations, with correction for atmospheric extinction (see Table~\ref{tab:obs}). Finally, the spectra from the three comet frames were transformed into the comet rest frame. Sky subtraction was not applied because the cometary [O~I] emission lines are well separated from the sky [O~I] features and are not affected by sky-background contamination (see the representative spectrum in Figure~\ref{fig:OI_lines}).

To isolate the cometary emission lines, we removed continuum components representing sunlight reflected by cometary dust grains and telluric absorption. The modeled continuum was constructed as the product of a high-resolution solar spectrum \citep{Meftah2023}, a wavelength-dependent dust reflectance spectrum, and a telluric transmittance spectrum. The dust reflectance spectrum was derived by dividing the continuum component of the reduced spectrum by the modeled solar spectrum, and the telluric transmittance spectrum was computed using the Planetary Spectrum Generator for the observing conditions at the time of the comet observations \citep{Villanueva2018}. 

After subtracting the modeled continuum and correcting for telluric absorption, we co-added the spectra from the three individual comet rest frames to produce the final gas-emission spectrum used in our analysis.

\begin{deluxetable*}{lccc cc}
\tablecaption{The measured intensities of the three [O I] emission lines, line velocity widths, the intensity ratio of the red doublet, and the green-to-red line ratio.\label{tab:oi_meas}}
\tablehead{
\colhead{} &
\multicolumn{3}{c}{\shortstack{Intensity [arb.\ units]\\(Observed line width [km s$^{-1}$])\\(Instrumental line width [km s$^{-1}$])\\(Intrinsic line width [km s$^{-1}$])}} &
\colhead{$I_{630.0}/I_{636.4}$} &
\colhead{$I_{557.7}/(I_{630.0}+I_{636.4})$} \\
\cline{2-4}
\colhead{} &
\colhead{[O I] $\lambda$557.7 nm} &
\colhead{[O I] $\lambda$630.0 nm} &
\colhead{[O I] $\lambda$636.4 nm} &
\colhead{} &
\colhead{}
}
\startdata
 & \shortstack{$7.85\pm0.59$\\(2.78$\pm$0.24)\\(1.71$\pm$0.16)\\(2.19$\pm$0.33)} &
   \shortstack{$17.24\pm0.46$\\(2.80$\pm$0.09)\\(1.73$\pm$0.06)\\(2.20$\pm$0.12)} &
   \shortstack{$5.92\pm0.36$\\(2.61$\pm$0.18)\\(1.78$\pm$0.14)\\(1.91$\pm$0.28)} &
   \shortstack{$2.91\pm0.20$\\~\\~\\~} &
   \shortstack{$0.339\pm0.027$\\~\\~\\~}
\enddata
\end{deluxetable*}

\section{Results and Discussion} \label{sec:style}
\subsection{[O~I] Line Measurements and Intensity Ratios} \label{subsec:OImeasurements}

Figure~\ref{fig:OI_lines} presents the extracted spectra around the three intrinsic cometary [O~I] lines at 557.7 nm, 630.0 nm, and 636.4 nm after removal of the dust-reflected continuum and telluric absorption. Owing to the large relative radial velocity of the comet with respect to the Earth during the observation ($\dot{\Delta} \sim 32.6$ km s$^{-1}$) compared with the instrumental resolution ($R = 72{,}000$; $\sim$4.2 km s$^{-1}$), the cometary [O~I] emission lines are well separated from the telluric [O~I] features. We measured the integrated line intensities by fitting single-Gaussian profiles to each [O~I] emission line, and the resulting line intensities and derived ratios are summarized in Table~\ref{tab:oi_meas}.

To assess potential optical-depth effects, we measured the intensity ratio of the [O~I] red doublet, $I_{630.0}/I_{636.4} = 2.91 \pm 0.20$. This value is consistent within the uncertainty with the optically thin theoretical branching ratio of $\sim$3.0 expected from the Einstein $A$ coefficients for the O($^1$D) to O($^3$P) transitions \citet{StoreyZeippen2000}. This agreement supports the conclusion that the measured red-doublet intensities are not significantly affected by optical-depth effects and that the relative line calibration is reliable at the level relevant for our analysis. The [O~I] green-to-red intensity ratio, defined as $G/R = I_{557.7}/(I_{630.0} + I_{636.4})$, is measured to be $0.339 \pm 0.027$. In Solar System comets, the [O~I] lines at 557.7 nm and 630.0 nm can be weakly contaminated by nearby C$_2$ and NH$_2$ emission lines, respectively (e.g., \citealt{ShinnakaKawakita2016}). We do not apply explicit corrections for these potential contaminations because we do not detect any isolated NH$_2$ or C$_2$ lines that would be expected to be stronger than the features adjacent to the [O~I] lines. Therefore, any residual blending is expected to be smaller than the uncertainties in the individual line intensities listed in Table~\ref{tab:oi_meas}.

\begin{figure}[t]
\centering
\includegraphics[width=1.0\columnwidth]{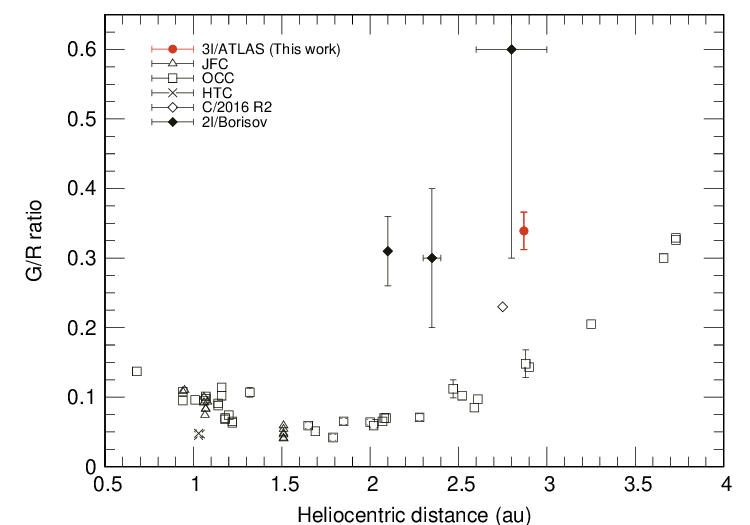}
\caption{The green-to-red intensity ratio of [O I] (G/R) in 3I (filled circle), compared with measurements for a sample of Solar System comets: Jupiter-family comets (JFCs; open triangles), Oort Cloud comets (OCCs; open squares), and Halley-type comets (HTCs; crosses) from \citet{Decock2013,McKay2015,ShinnakaKawakita2016,Shinnaka2020}, as well as C/2016 R2 from \citet{Opitom2019} and 2I/Borisov from \citet{Opitom2021}.}
\label{fig:GR_vs_rh}
\end{figure}

Figure~\ref{fig:GR_vs_rh} compares the measured $G/R$ ratio of 3I at $r_{\mathrm{h}} = 2.87$ au with published $G/R$ measurements for Solar System comets spanning Jupiter-family comets (JFCs), Oort Cloud comets (OCCs), and Halley-type comets (HTCs), as well as the CO-rich comet C/2016 R2 and the interstellar comet 2I/Borisov \citep{Decock2013,McKay2015,ShinnakaKawakita2016,Shinnaka2020,Opitom2019, Opitom2021}. Apart from the peculiar CO-dominated coma of C/2016 R2, Solar System comets in these samples tend to occupy a similar range of $G/R$ ratios at a given heliocentric distance. At $r_{\mathrm{h}} = 2.9$ au, 3I exhibits a comparatively elevated $G/R$ ratio relative to many Solar System comets, whereas 2I/Borisov shows values comparable to, or in some cases exceeding, those measured for 3I. These comparisons provide an observational context for interpreting the oxygen-line ratios in 3I within the broader comet population.

We also measured the intrinsic line velocity widths of the [O~I] lines, expressed as the full width at half maximum (FWHM) in km s$^{-1}$. The observed line profiles are the convolution of the intrinsic cometary line shape with the instrumental profile. The instrumental profile was estimated from unblended Th--Ar lamp lines at each [O~I] wavelength, and the intrinsic widths were then derived using Equations~(7) and (9) of \citet{Decock2013}. We found that the FWHM of the green line is comparable to that of the red doublet within the uncertainties, consistent with trends reported for several Solar System comets \citep{Decock2013}. The resulting FWHM values are also listed in Table~\ref{tab:oi_meas}.

\begin{deluxetable*}{lccc}
\tablecaption{Photodissociation Rates at 1 au and the Branching Ratio Used to Convert an Observed Green-to-Red Line Ratio to a CO$_2$/H$_2$O Abundance Ratio\label{tab:gr_conversion}}
\tablewidth{0pt}
\tablehead{
\colhead{} & \colhead{Case A} & \colhead{Case B} & \colhead{Case C}
}
\startdata
$W_{\mathrm{H_2O}}^{\mathrm{green}}$ & $3.20 \times 10^{-8}$ s$^{-1}$\tablenotemark{a} & $2.28 \times 10^{-8}$ s$^{-1}$\tablenotemark{b} & $6.40 \times 10^{-8}$ s$^{-1}$\tablenotemark{c} \\
$W_{\mathrm{H_2O}}^{\mathrm{red}}$   & $8.00 \times 10^{-7}$ s$^{-1}$\tablenotemark{a} & $5.38 \times 10^{-7}$ s$^{-1}$\tablenotemark{b} & $8.44 \times 10^{-7}$ s$^{-1}$\tablenotemark{c} \\
$W_{\mathrm{CO_2}}^{\mathrm{green}}$ & $7.20 \times 10^{-7}$ s$^{-1}$\tablenotemark{a} & $7.20 \times 10^{-7}$ s$^{-1}$\tablenotemark{a} & $3.30 \times 10^{-7}$ s$^{-1}$\tablenotemark{c} \\
$W_{\mathrm{CO_2}}^{\mathrm{red}}$   & $5.25 \times 10^{-7}$ s$^{-1}$\tablenotemark{a} & $5.25 \times 10^{-7}$ s$^{-1}$\tablenotemark{a} & $4.95 \times 10^{-7}$ s$^{-1}$\tablenotemark{c} \\
$\beta_{\mathrm{green}}$             & 0.91\tablenotemark{d} & 0.91\tablenotemark{d} & 0.91\tablenotemark{d} \\
CO$_2$/H$_2$O ratio                  & $0.507 \pm 0.048$ & $0.339 \pm 0.032$ & $2.12 \pm 0.27$ \\
\enddata
\tablenotetext{a}{\citet{RaghuramBhardwaj2013} for the quiet-Sun case, converted to 1 au.}
\tablenotetext{b}{\citet{Kawakita2022} for the quiet-Sun case.}
\tablenotetext{c}{Values of McKay2015B in Table~6 of \citet{McKay2016}.}
\tablenotetext{d}{\citet{Slanger2006}.}
\end{deluxetable*}

\subsection{CO$_{2}$/H$_{2}$O ratio inferred from the [O I] green-to-red ratio} \label{subsec:CO$_{2}$H$_{2}$O}

Independent of direct infrared detections of CO$_2$ and H$_2$O, the [O~I] intensity ratios can be used to estimate the relative importance of oxygen-bearing parent species in cometary comae. Following Equation~(12) of \citet{Decock2013}, the CO$_2$/H$_2$O abundance ratio can be expressed in terms of the observed $G/R$ ratio and the effective photodissociation rates that produce O($^1$S) and O($^1$D):

\begin{equation}
\frac{N_{\mathrm{CO_2}}}{N_{\mathrm{H_2O}}}
=
\frac{(G/R)\,W_{\mathrm{H_2O}}^{\mathrm{red}}-\beta^{\mathrm{green}}\,W_{\mathrm{H_2O}}^{\mathrm{green}}}
{\beta^{\mathrm{green}}\,W_{\mathrm{CO_2}}^{\mathrm{green}}-(G/R)\,W_{\mathrm{CO_2}}^{\mathrm{red}}}
\end{equation}

Here, $N_{\mathrm{H_2O}}$ and $N_{\mathrm{CO_2}}$ are the column densities of H$_2$O and CO$_2$, respectively. $W_{\mathrm{H_2O}}^{\mathrm{green}}$ and $W_{\mathrm{H_2O}}^{\mathrm{red}}$ ($W_{\mathrm{CO_2}}^{\mathrm{green}}$ and $W_{\mathrm{CO_2}}^{\mathrm{red}}$) denote the effective photodissociation rates populating O($^1$S) and O($^1$D), respectively, for each parent molecule. $\beta_{\mathrm{green}}$ is the branching ratio for production of the [O~I] line at 557.7 nm from O($^1$S). This conversion assumes that the production of O($^1$S) and O($^1$D) atoms in the coma is dominated by H$_2$O and CO$_2$. Although other oxygen-bearing species (e.g., CO and O$_2$) may contribute to the production of atomic oxygen, their influence on the $G/R$ ratio is expected to be small. For example, under typical conditions, \citep{Decock2013} showed that including CO as an additional parent changes the $G/R$ ratio only weakly (from 0.08 to 0.10) for CO abundances of 10\%--80\%.

\begin{deluxetable*}{lcccl}
\tablecaption{Summary of CO$_2$/H$_2$O Ratios of 3I/ATLAS from Space-Based Observations and Our [O~I]-Based Constraint\label{tab:co2h2o_summary}}
\tablewidth{0pt}
\tablehead{
\colhead{Date} &
\colhead{$r_{\mathrm{h}}$ (au)} &
\colhead{CO$_2$/H$_2$O} &
\colhead{Telescope/Instrument} &
\colhead{Reference}
}
\startdata
2025 August 6      & 3.32      & $7.6 \pm 0.3$   & JWST/NIRSpec & \citet{Cordiner2025} \\
2025 August 8--12  & 3.3--3.1  & $>4.4$          & SPHEREx      & \citet{Lisse2025} \\
2025 December 8--15& 2.0--2.2  & $0.21 \pm 0.05$ & SPHEREx      & \citet{Lisse2026} \\
2025 December 15--16 & 2.19    & $3.16 \pm 0.06$ & JWST/MIRI    & \citet{Belyakov2026} \\
2025 December 27   & 2.54      & $7.28 \pm 0.17$ & JWST/MIRI    & \citet{Belyakov2026} \\
2026 January 7     & 2.87      & 0.307--2.39     & Subaru/HDS   & This work \\
\enddata
\tablenotetext{}{Unless otherwise noted, uncertainties and limits in this table are quoted at the 1$\sigma$ level. For our [O~I]-based result, the CO$_2$/H$_2$O value is reported as the range spanned by Cases A--C. For the SPHEREx entry on 2025 August 8--12 (C.~Lisse et al.\ 2025), $Q(\mathrm{H_2O})$ was a 3$\sigma$ upper limit, and the corresponding CO$_2$/H$_2$O value is therefore given as a lower limit. Perihelion occurred on UT 2025 October 29.}
\end{deluxetable*}

To convert our measured $G/R$ ratio into the CO$_2$/H$_2$O ratio, we adopt three parameter sets that are commonly used in the literature (see Table~\ref{tab:gr_conversion}). In Case A, we use the theoretical rates for both H$_2$O and CO$_2$ from \citet{RaghuramBhardwaj2013} for the quiet-Sun case, scaled to 1 au. In Case B, we use the CO$_2$ rates from \citet{RaghuramBhardwaj2013} and the H$_2$O rates from \citet{Kawakita2022}. In Case C, we use the empirical rates for both H$_2$O and CO$_2$ reported in Table~6 of \citet{McKay2016}. Applying these parameter sets to the measured $G/R$ ratio yields CO$_2$/H$_2$O $= 0.507 \pm 0.048$ for Case A, CO$_2$/H$_2$O $= 0.339 \pm 0.032$ for Case B, and CO$_2$/H$_2$O $= 2.12 \pm 0.27$ for Case C. The quoted uncertainties reflect propagation of the 1$\sigma$ measurement error on the $G/R$ ratio. Recomputing the $G/R$ ratio with active-Sun photodissociation rates yields values unchanged within our quoted precision, indicating that our inferred CO$_2$/H$_2$O ratio is not sensitive to the assumed solar activity level for the purposes of this work. We emphasize that the absolute CO$_2$/H$_2$O ratio inferred from $G/R$ is model dependent, as illustrated by the difference among Cases A--C. Nevertheless, all three cases indicate that 3I has a higher CO$_2$/H$_2$O ratio than is typical for Solar System comets at similar heliocentric distances.

We interpret our result by comparing it with published infrared measurements of 3I/ATLAS obtained at different epochs and heliocentric distances. For clarity, we compile the key published infrared constraints in Table~4. On the inbound trajectory near $r_{\mathrm{h}} \sim 3.3$ au, JWST/NIRSpec IFU observations obtained on 2025 August 6 at $r_{\mathrm{h}} = 3.32$ au reported extremely CO$_2$-rich outgassing, with CO$_2$/H$_2$O $= 7.6 \pm 0.3$ and CO/H$_2$O $= 1.65 \pm 0.09$ \citep{Cordiner2025}. SPHEREx imaging spectrophotometric observations obtained on 2025 August 8--12, spanning $r_{\mathrm{h}} = 3.3$--$3.1$ au, yielded CO$_2$/H$_2$O $> 4.4$ and CO/H$_2$O $> 1.9$ \citep{Lisse2025}. After perihelion, Ly$\alpha$ measurements indicate that the H$_2$O production increased dramatically, reaching $Q(\mathrm{H_2O}) = (3.17 \pm 0.14) \times 10^{29}$ s$^{-1}$ near perihelion and then declining with increasing heliocentric distance \citep{Combi2026}.

By December 2025 on the outbound trajectory at $r_{\mathrm{h}} = 2.0$--$2.2$ au (2025 December 8--15), SPHEREx reported $Q(\mathrm{H_2O}) = (1.4 \pm 0.28) \times 10^{28}$ s$^{-1}$ and CO$_2$/H$_2$O $= 0.21 \pm 0.05$ \citep{Lisse2026}, which is more typical of Solar System comets at comparable heliocentric distance \citep[e.g.,][]{Ootsubo2012}. JWST/MIRI also obtained two post-perihelion epochs in December 2025 and derived substantially higher CO$_2$/H$_2$O ratios of $3.16 \pm 0.06$ at $r_{\mathrm{h}} = 2.19$ au on 2025 December 15--16 and $7.28 \pm 0.17$ at $r_{\mathrm{h}} = 2.54$ au on 2025 December 27 \citep{Belyakov2026}.

The large discrepancy between the outbound CO$_2$/H$_2$O ratios reported by JWST and SPHEREx may reflect differences in the spatial scales sampled by the observations. We note, however, that the infrared measurements are direct probes of H$_2$O and CO$_2$, and are therefore less model dependent than the [O~I]-based estimates. JWST employs a relatively small field of view, so extended coma emission at large projected nucleocentric distances may not be fully captured. If H$_2$O has a more extended spatial distribution than CO$_2$, then a limited aperture can underestimate H$_2$O and bias CO$_2$/H$_2$O toward higher values. As noted by \citet{Belyakov2026}, while efforts were made to account for extended H$_2$O sources, the limited field of view of the JWST/MIRI observations may still introduce uncertainty in these ratios. At the same time, the [O~I] lines are forbidden emission with short radiative lifetimes ($\sim$0.74 s for O($^1$S) and $\sim$110 s for O($^1$D); \citealt{Kramida2024}). Therefore, the production and decay can occur over relatively small spatial scales, potentially within a single observational aperture. Temporal variability and model-dependent choices in the [O~I] conversion and/or infrared retrievals may also contribute to the observed dispersion among reported mixing ratios.

Table~\ref{tab:co2h2o_summary} summarizes the CO$_2$/H$_2$O ratio of interstellar comet 3I as a function of heliocentric distance by combining published infrared measurements with our [O~I]-based constraints. Our result at $r_{\mathrm{h}} = 2.87$ au provides an additional post-perihelion constraint at a heliocentric distance distinct from those sampled by the outbound SPHEREx and JWST/MIRI observations. If the CO$_2$/H$_2$O ratio indeed changes substantially across perihelion, these measurements support the scenario proposed by \citet{Lisse2026}, in which 3I evolved from an extremely CO$_2$-rich composition on the inbound trajectory to more H$_2$O-dominated outgassing on the outbound trajectory. Because the available post-perihelion measurements were obtained at different heliocentric distances and with different instrumental apertures, they may not probe the same coma region or sample the same active region on the nucleus responsible for the outgassing. One plausible, though not unique, explanation is depth-dependent radial heterogeneity, in which CO$_2$- and CO-enriched near-surface layers dominate the pre-perihelion activity, while deeper, more H$_2$O-rich material contributes more strongly after perihelion. Additional observations of 3I at different heliocentric distances, together with thermophysical and compositional modeling, will be needed to clarify the origin of the observed pre- and post-perihelion differences in volatile composition.

\section{Summary and Conclusions}
We present high-resolution optical spectroscopy of interstellar comet 3I/ATLAS (C/2025 N1) obtained with Subaru/HDS on UT 2026 January 7 at $r_{\mathrm{h}} = 2.87$ au on the outbound trajectory. We measure the [O~I] green-to-red intensity ratio of $G/R = 0.339 \pm 0.027$, which is higher than values for most Solar System comets at similar heliocentric distances, but comparable to that of 2I/Borisov.

Using the standard conversion from $G/R$ to CO$_2$/H$_2$O, we infer the CO$_2$/H$_2$O abundance ratio under the assumption that H$_2$O and CO$_2$ dominate the production of O($^1$S) and O($^1$D). Applying three parameter sets yields CO$_2$/H$_2$O $= 0.507 \pm 0.048$ (Case A), $0.339 \pm 0.032$ (Case B), and $2.12 \pm 0.27$ (Case C), demonstrating that the absolute abundance ratio derived from $G/R$ is model dependent. Nevertheless, all cases indicate that 3I's post-perihelion CO$_2$/H$_2$O ratio at $r_{\mathrm{h}} = 2.87$ au lies above the typical range for Solar System comets at comparable heliocentric distances.

Finally, we place our [O~I]-based constraint in the context of published infrared measurements obtained on both the inbound and outbound trajectories. Our measurement provides an independent post-perihelion constraint on the CO$_2$/H$_2$O ratio of 3I/ATLAS, complementing the space-based infrared results and helping to assess compositional evolution across perihelion, which may include a decrease in the CO$_2$/H$_2$O ratio from the inbound to the outbound trajectory.

\begin{acknowledgments}
The authors thank the anonymous referee for helpful suggestions that improved the manuscript. This research is based on data collected at the Subaru Telescope, operated by the National Astronomical Observatory of Japan (NAOJ). We are honored and grateful for the opportunity to observe the Universe from Maunakea, which has cultural, historical, and natural significance in Hawaii. This research was supported by Japan Society for the Promotion of Science (JSPS) Grant Numbers JP20K14541 (YS), JP21H04498 (HKa), JP23K25930 (TO), and JP25Kxxxxx (HKa). This research was partly supported by Koyama Space Science Institute of Kyoto Sangyo University.
\end{acknowledgments}

\facilities{Subaru Telescope}


\begin{thebibliography}{}

\bibitem[M. Belyakov et al.(2026)]{Belyakov2026}
Belyakov, M., Wong, I., Bolin, B. T., et al. 2026, arXiv e-prints, arXiv:2601.22034, doi:10.48550/arXiv.2601.22034

\bibitem[N. Biver et al.(2024)]{Biver2024}
Biver, N., Dello Russo, N., Opitom, C., Rubin, M., \& Dotson, R. 2024, in \textit{Comets III}, ed. K. J. Meech, M. R. Combi, D. Bockel\'ee-Morvan, S. N. Raymond, \& M. E. Zolensky (University of Arizona Press), 459--498

\bibitem[M. Combi et al.(2026)]{Combi2026}
Combi, M. R., M\"akinen, T., Bertaux, J.-L., et al. 2026, \apj, 996, 17, doi:10.3847/2041-8213/ae3bd8

\bibitem[M. Cordiner et al.(2025)]{Cordiner2025}
Cordiner, M., Roth, N. X., Kelley, M. S. P., et al. 2025, \apjl, 991, L43, doi:10.3847/2041-8213/ae0647

\bibitem[A. Decock et al.(2013)]{Decock2013}
Decock, A., Jehin, E., Hutsem\'ekers, D., \& Manfroid, J. 2013, \aap, 555, A34, doi:10.1051/0004-6361/201219154

\bibitem[A. Decock et al.(2015)]{Decock2015}
Decock, A., Jehin, E., Rousselot, P., et al. 2015, \aap, 573, A1, doi:10.1051/0004-6361/201424403

\bibitem[L. Denneau et al.(2025)]{Denneau2025}
Denneau, L., Siverd, R., Tonry, J., et al. 2025, Minor Planet Electronic Circulars, 2025-N12

\bibitem[M. Festou \& P. Feldman(1981)]{FestouFeldman1981}
Festou, M. C., \& Feldman, P. D. 1981, \aap, 103, 154

\bibitem[T. Frincke et al.(2026)]{Frincke2026}
Frincke, T. T., Yaginuma, A., Noonan, J. W., et al. 2026, \mnras, 545, staf1994, doi:10.1093/mnras/staf1994

\bibitem[R. Furusho et al.(2006)]{Furusho2006}
Furusho, R., Kawakita, H., Fuse, T., Watanabe, J., et al. 2006, ASR, 38, 1983, doi:10.1016/j.asr.2006.06.003

\bibitem[E. Jehin et al.(2025)]{Jehin2025}
Jehin, E., Hmiddouch, S., Aravind, D., et al. 2025, The Astronomer's Telegram, 17515

\bibitem[H. Kawakita(2022)]{Kawakita2022}
Kawakita, H. 2022, \apj, 931, 24, doi:10.3847/1538-4357/ac67e2

\bibitem[A. Kramida et al.(2024)]{Kramida2024}
Kramida, A., Ralchenko, Yu., Reader, J., \& NIST ASD Team. 2024, NIST Atomic Spectra Database (ver. 5.12) (Gaithersburg, MD: National Institute of Standards and Technology), doi:10.18434/T4W30F

\bibitem[C. Lisse et al.(2025)]{Lisse2025}
Lisse, C. M., Bach, Y. P., Bryan, S., et al. 2025, RNAAS, 9, 242, doi:10.3847/2515-5172/ae0293

\bibitem[C. Lisse et al.(2026)]{Lisse2026}
Lisse, C. M., Bach, Y. P., Bryan, S. A., et al. 2026, RNAAS, 10, 26, doi:10.3847/2515-5172/ae3f95

\bibitem[R. Maggiolo et al.(2026)]{Maggiolo2026}
Maggiolo, R., Dhooghe, F., Gronoff, G. P., de Keyser, J., \& Cessateur, G. 2026, \apjl, 996, L34, doi:10.3847/2041-8213/ae2fff

\bibitem[M. Meftah et al.(2023)]{Meftah2023}
Meftah, M., Sarkissian, A., Keckhut, P., \& Hauchecorne, A. 2023, Remote Sensing, 15, 3560, doi:10.3390/rs15143560

\bibitem[A. McKay et al.(2015)]{McKay2015}
McKay, A. J., Cochran, A. L., DiSanti, M. A., et al. 2015, \icarus, 250, 504, doi:10.1016/j.icarus.2014.12.023

\bibitem[A. McKay et al.(2016)]{McKay2016}
McKay, A. J., Kelley, M. S. P., Cochran, A. L., et al. 2016, \icarus, 266, 249, doi:10.1016/j.icarus.2015.11.004

\bibitem[K. Noguchi et al.(2002)]{Noguchi2002}
Noguchi, K., Aoki, W., Kawanomoto, S., et al. 2002, \pasj, 54, 855, doi:10.1093/pasj/54.6.855

\bibitem[T. Ootsubo et al.(2012)]{Ootsubo2012}
Ootsubo, T., Kawakita, H., Hamada, S., et al. 2012, \apj, 752, 15, doi:10.1088/0004-637X/752/1/15

\bibitem[C. Opitom et al.(2019)]{Opitom2019}
Opitom, C., Hutsem\'ekers, D., Jehin, E., et al. 2019, \aap, 624, A64, doi:10.1051/0004-6361/202142829

\bibitem[C. Opitom et al.(2021)]{Opitom2021}
Opitom, C., Jehin, E., Hutsem\'ekers, D., et al. 2021, \aap, 650, L19, doi:10.1051/0004-6361/202141245

\bibitem[C. Opitom et al.(2025)]{Opitom2025}
Opitom, C., Snodgrass, C., Jehin, E., et al. 2025, \mnras, 544, L31, doi:10.1093/mnrasl/slaf095

\bibitem[S. Raghuram \& A. Bhardwaj(2013)]{RaghuramBhardwaj2013}
Raghuram, S., \& Bhardwaj, A. 2013, \icarus, 223, 91, doi:10.1016/j.icarus.2012.11.032

\bibitem[R. Rahatgaonkar et al.(2025)]{Rahatgaonkar2025}
Rahatgaonkar, R., Carvajal, J. P., Puzia, T. H., et al. 2025, \apjl, 995, L34, doi:10.3847/2041-8213/ae1cbc

\bibitem[L. Salazar Manzano et al.(2025)]{SalazarManzano2025}
Salazar Manzano, L. E., Lin, H. W., Taylor, A. G., et al. 2025, \apjl, 993, L23, doi:10.3847/2041-8213/ae1232

\bibitem[Y. Shinnaka et al.(2020)]{Shinnaka2020}
Shinnaka, Y., Kawakita, H., \& Tajitsu, A. 2020, \aj, 159, 203, doi:10.3847/1538-3881/ab7d34

\bibitem[Y. Shinnaka \& H. Kawakita(2016)]{ShinnakaKawakita2016}
Shinnaka, Y., \& Kawakita, H. 2016, \aj, 152, 145, doi:10.3847/0004-6256/152/5/145

\bibitem[T. Slanger et al.(2006)]{Slanger2006}
Slanger, T. G., Cosby, P. C., Sharpee, B. D., Minschwaner, K. R., \& Siskind, D. E. 2006, Journal of Geophysical Research (Space Physics), 111, A12318, doi:10.1029/2006JA011972

\bibitem[P. Storey \& C. Zeippen(2000)]{StoreyZeippen2000}
Storey, P. J., \& Zeippen, C. J. 2000, \mnras, 312, 813, doi:10.1046/j.1365-8711.2000.03184.x

\bibitem[B. Yang et al.(2025)]{Yang2025}
Yang, B., Meech, K. J., Connelley, M., Zhao, R., \& Keane, J. V. 2025, \apjl, 992, L9, doi:10.3847/2041-8213/ae08a7

\bibitem[G. Villanueva et al.(2018)]{Villanueva2018}
Villanueva, G. L., Smith, M. D., Protopapa, S., Faggi, S., \& Mandell, A. M. 2018, Journal of Quantitative Spectroscopy and Radiative Transfer, 217, 86, doi:10.1016/j.jqsrt.2018.05.023

\end{thebibliography}
\end{document}